\documentclass[12pt]{article}
\usepackage{graphicx}
\usepackage{xspace}

\newcommand\pubnumber{MCNET-17-19}
\newcommand\pubdate{\today}

\def\institute{Institute for Theoretical Physics,\\
  Georg-August-Univesity G\"ottingen,\\
  Friedrich-Hund-Platz 1,\\37077 G\"ottingen, GERMANY}
\def\support{\footnote{Work supported by the European Union as part of the
    H2020 Marie Sklodowska-Curie Initial Training Network MCnetITN3 (722104)
    and BMBF under contract 05H15MGCAA.}}

\def\Title#1{\begin{center} {\Large #1 } \end{center}}
\def\Author#1{\begin{center}{ \sc #1} \end{center}}
\def\Address#1{\begin{center}{ \it #1} \end{center}}

\newcommand\pubblock{\rightline{\begin{tabular}{l} \pubnumber\\
         \pubdate  \end{tabular}}}
\newenvironment{Abstract}{\begin{quotation}  }{\end{quotation}}
\newenvironment{Presented}{\begin{quotation} \begin{center} 
             PRESENTED AT\end{center}\bigskip 
      \begin{center}\begin{large}}{\end{large}\end{center} \end{quotation}}

\newcommand{\Recola}{{\sc Recola}\xspace}
\newcommand{\Sherpa}{{\sc Sherpa}\xspace}

\newcommand{\OpenLoops}{O\protect\scalebox{0.8}{PEN}L\protect\scalebox{0.8}{OOPS}\xspace}

\def\beq{\begin{equation}}
\def\eeq#1{\label{#1}\end{equation}}
\def\eeqn{\end{equation}}

\def\beqa{\begin{eqnarray}}
\def\eeqa#1{\label{#1}\end{eqnarray}}
\def\eeqan{\end{eqnarray}}

\let\bar=\overbar

\def\Dslash{\not{\hbox{\kern-4pt $D$}}}
\def\dslash{\not{\hbox{\kern-2pt $\del$}}}

\def\msb{{\bar{\ssstyle M \kern -1pt S}}}

\begin{document}
\begin{titlepage}
\pubblock

\vfill
\Title{Automated QCD and Electroweak Corrections with Sherpa}
\vfill
\Author{ Steffen Schumann\support}
\Address{\institute}
\vfill
\begin{Abstract}
  Precise theoretical predictions are vital for the interpretation of
  Standard Model measurements and facilitate conclusive searches
  for New Physics phenomena at the LHC. In this contribution I highlight
  some of the ongoing efforts in the framework of the \Sherpa event generator
  to provide accurate and realistic simulations of Standard Model production
  processes. This includes the automated evaluation of NLO QCD and NLO EW
  corrections and their consistent consideration in the parton-shower
  evolution of scattering events. Particular emphasis is given to examples
  and applications involving the production of top quarks. 
\end{Abstract}
\vfill
\begin{Presented}
$10^{th}$ International Workshop on Top Quark Physics\\
Braga, Portugal,  September 18--22, 2017
\end{Presented}
\vfill
\end{titlepage}
\def\thefootnote{\fnsymbol{footnote}}
\setcounter{footnote}{0}

\section{Introduction}

Precise predictions for the production rates and kinematical distributions
of Standard Model (SM) processes are of utmost importance for measurements of
SM parameters, {\it e.g.} couplings and particle quantum numbers, and meaningful
searches for New Physics phenomena. To link accurate perturbative calculations
to particle-level final states in collider experiments Monte Carlo event generators
prove to be indispensable. Ideally these tools should provide the highest
possible perturbative accuracy for the considered hard-production
process, consistently combined with a resummation of large scale logarithms
and linked to phenomenological models for non-pertubative aspects, such as
hadronization and underlying event \cite{Buckley:2011ms}. This should be
achieved such, that the perturbative accuracy is preserved and that related
theoretical uncertainties remain quantifiable, allowing to reliably contrast
experimental measurements with SM expectations. 

In this short contribution I highlight some recent achievements and results
from the development of the \Sherpa event-generation framework \cite{Gleisberg:2008ta},
targeted towards its application in top-quark physics at the LHC.

\section{Fixed-order Matrix Elements}

Central to accurate predictions for SM production rates, based on fixed-order
perturbation theory, is the inclusion of higher-order corrections. Given that a
huge variety of processes needs to be considered, for the dominant NLO QCD
contributions these calculations have largely been automated in the framework
of parton-level event generators. Within \Sherpa automation is achieved through
efficient tree-level matrix-element generators that also implement the construction
of suitable infrared-subtraction terms in the Catani--Seyour dipole
approach~\cite{Gleisberg:2007md}. For the evaluation of the one-loop amplitude
contributions interfaces to dedicated codes such as \OpenLoops~\cite{Cascioli:2011va}
and \Recola~\cite{Actis:2016mpe} are employed. It remains the challenge to obtain
results for very complex final states, in particular when featuring several QCD partons. 

%%%%%%%%%%%%%%%%%%%%%%%%%%%%%%%%%%%%%%%%%%%%%%%%%%%%%%%%%%%%%%%%%%%%%%%%%%%%%%%%%%%%%%%%%%%%
%%%%%%%%%%%%%%%%%%%%%%%%%%%%%%%%%%%%%%%%%%%%%%%%%%%%%%%%%%%%%%%%%%%%%%%%%%%%%%%%%%%%%%%%%%%%

\begin{figure}[h!]
\begin{minipage}{0.49\textwidth}
\begin{center}
  \includegraphics[scale=0.525,trim=0 25 10 0,clip]{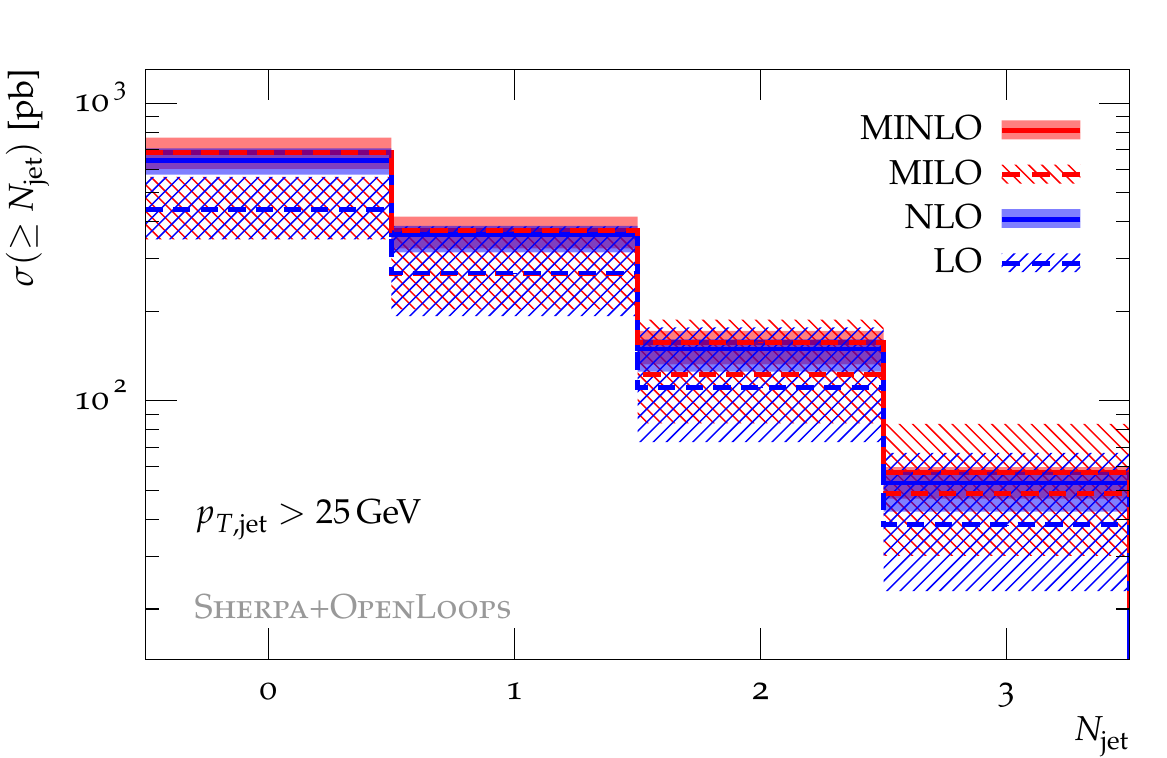}\\[-4pt]
  \includegraphics[scale=0.525,trim=0 25 10 0,clip]{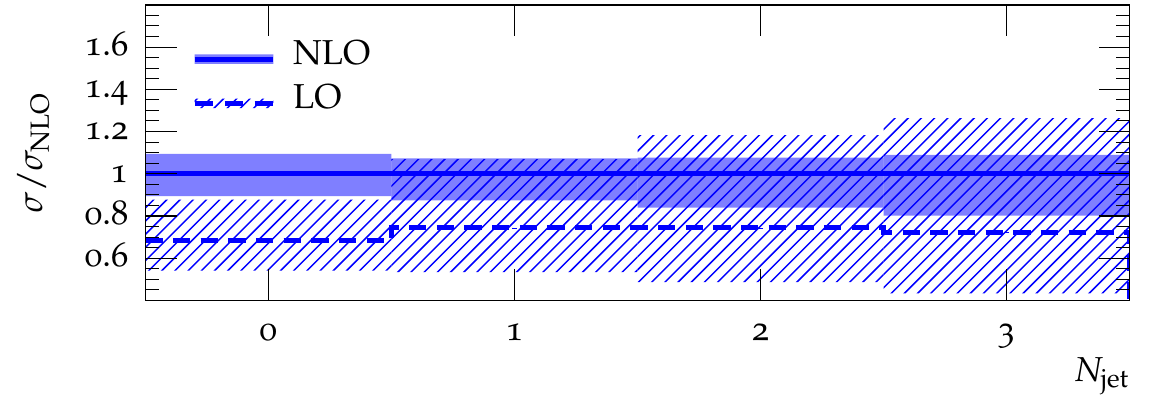}\\[-4pt]
  \includegraphics[scale=0.525,trim=0 25 10 0,clip]{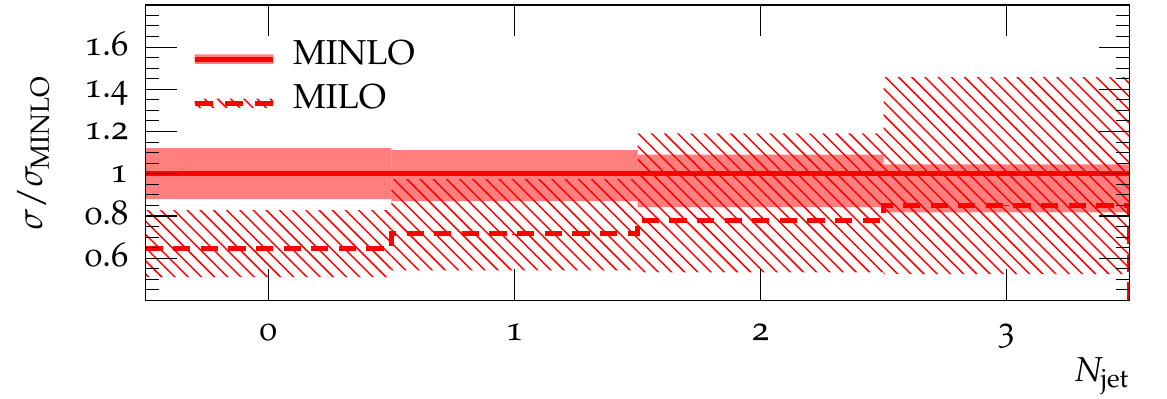}\\[-4pt]
  \includegraphics[scale=0.525,trim=0 0 10 0,clip]{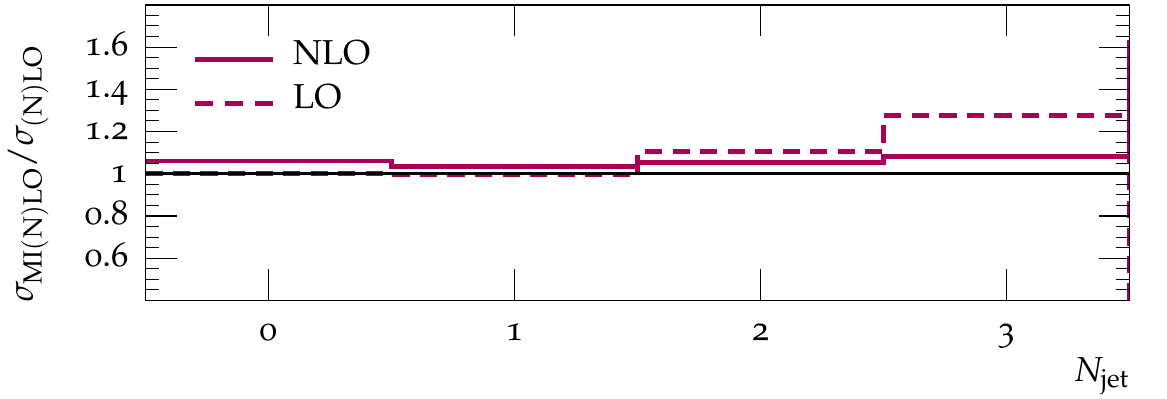}
\end{center}
\end{minipage}
\begin{minipage}{0.49\textwidth}
  \begin{center}
    \includegraphics[scale=0.425,trim=0 25 10 0,clip]{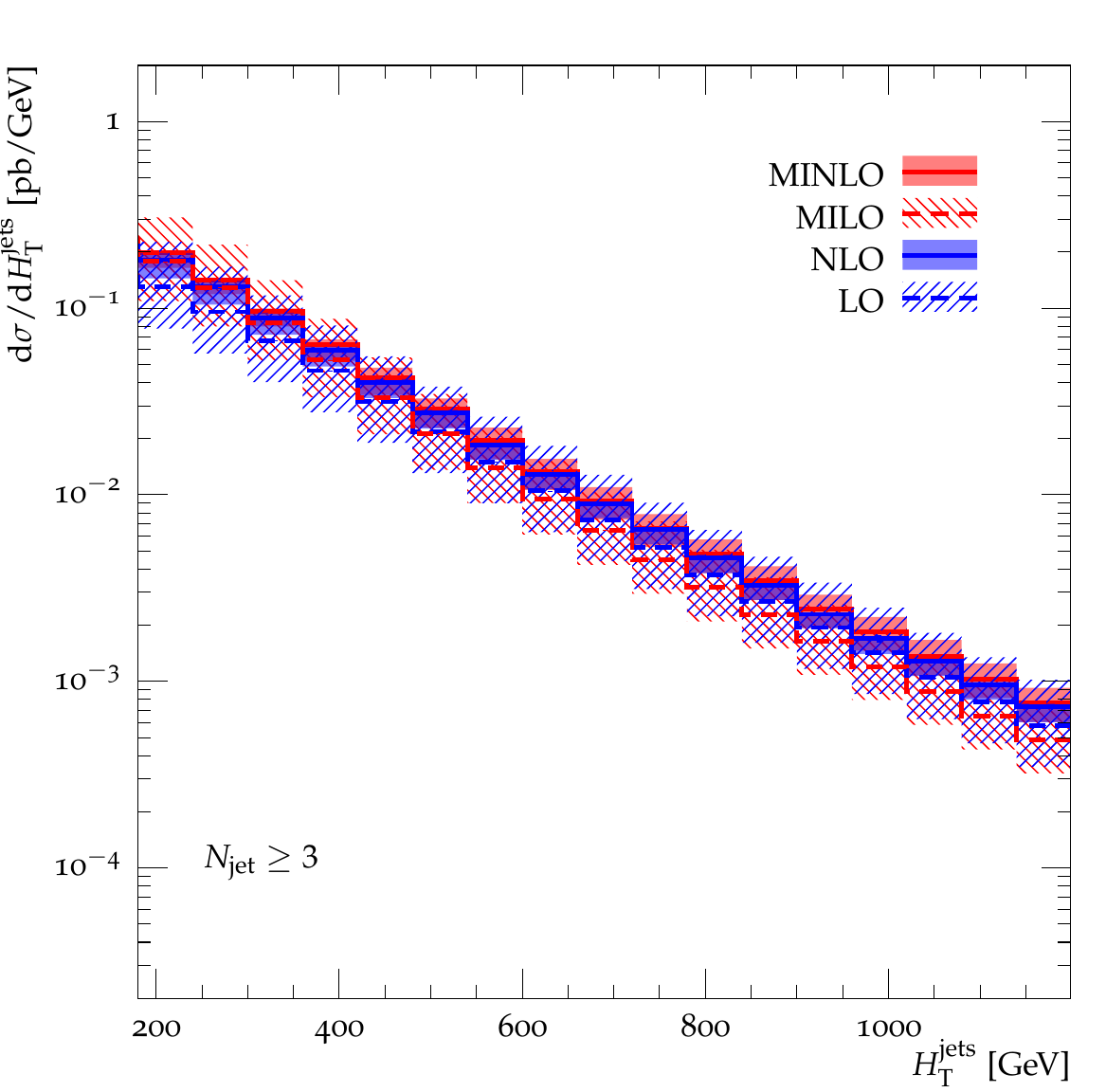}\\[-3.675pt]
    \includegraphics[scale=0.425,trim=0 25 10 0,clip]{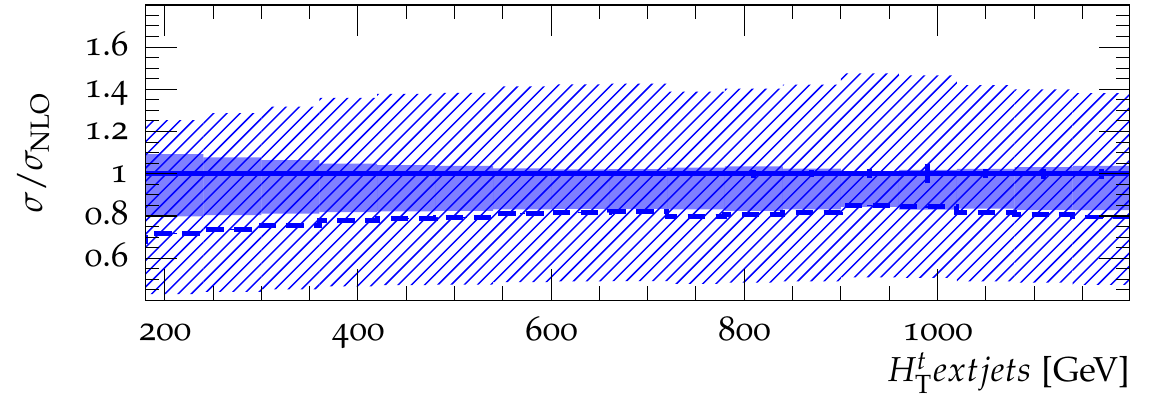}\\[-3.675pt]
    \includegraphics[scale=0.425,trim=0 25 10 0,clip]{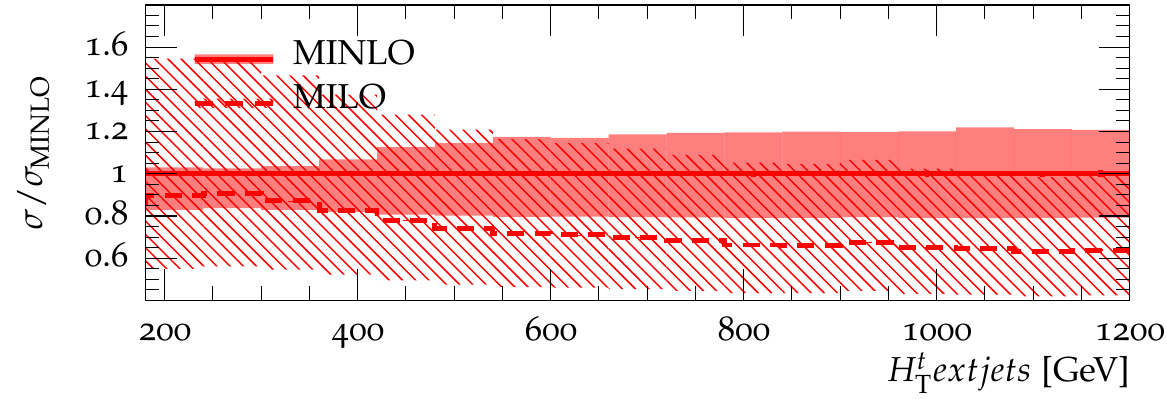}\\[-3.675pt]
    \includegraphics[scale=0.425,trim=0 0 10 0,clip]{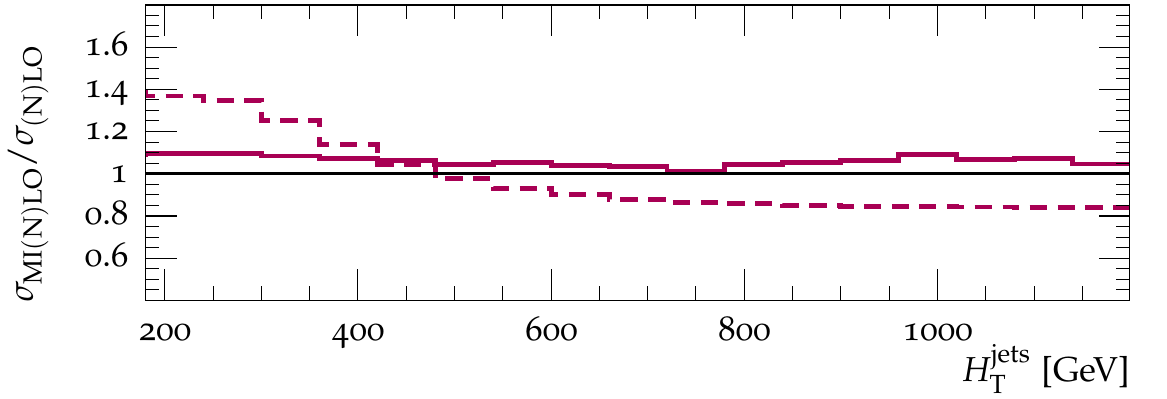}
  \end{center}
\end{minipage}
\caption{
Inclusive $t\bar{t}+$jets production at $\sqrt{s}=13$ TeV with a minimum number 
$N_{\rm jet}=0,1,2,3$ of jets with $p_{T,j}\ge 25\,$GeV (left panel).
Distribution of the total light-jet transverse energy for $pp\to t\bar{t}+3$\,jets production
with $p_{T,j}\ge 25$\,GeV (right panel). Figures taken from Ref.~\cite{Hoche:2016elu}.}
\label{fig:tt3jets}
\end{figure}

%%%%%%%%%%%%%%%%%%%%%%%%%%%%%%%%%%%%%%%%%%%%%%%%%%%%%%%%%%%%%%%%%%%%%%%%%%%%%%%%%%%%%%%%%%%%
%%%%%%%%%%%%%%%%%%%%%%%%%%%%%%%%%%%%%%%%%%%%%%%%%%%%%%%%%%%%%%%%%%%%%%%%%%%%%%%%%%%%%%%%%%%%

The first evaluation of NLO QCD corrections to $t\bar{t}+3$jets at the
LHC have been presented in \cite{Hoche:2016elu}. In Fig.~\ref{fig:tt3jets} predictions
for the jet-multiplicity and the $H_T^{\rm jets}$ distribution in LHC
collisions at $\sqrt{s}=13$ TeV are shown. The NLO QCD results, evaluated for
$\mu_{R/F}=H_T/2$, are compared to the corresponding LO estimates. Furthermore,
results obtained using the Multi-scale Improved scale setting prescription at LO and NLO,
dubbed MILO and MINLO, respectively, are presented. In both scale-setting schemes the
uncertainty estimates, {\it i.e.} 7-point scale variations, indicated by the bands,
reduce significantly at NLO. While for both observables the QCD corrections are rather
flat for the scale choice $\mu_{R/F}=H_T/2$, in the MI scheme a shape distortion
is observed. However, the NLO and MINLO results agree within $10\%$. 

%%%%%%%%%%%%%%%%%%%%%%%%%%%%%%%%%%%%%%%%%%%%%%%%%%%%%%%%%%%%%%%%%%%%%%%%%%%%%%%%%%%%%%%%%%%%
%%%%%%%%%%%%%%%%%%%%%%%%%%%%%%%%%%%%%%%%%%%%%%%%%%%%%%%%%%%%%%%%%%%%%%%%%%%%%%%%%%%%%%%%%%%%

With the increasing precision of the LHC measurements, the inclusion of NLO electroweak (EW)
corrections also becomes very relevant. Accordingly, there are strong ongoing efforts to
extend the above mentioned methods for the automated evaluation of NLO corrections
to the EW sector.    

%%%%%%%%%%%%%%%%%%%%%%%%%%%%%%%%%%%%%%%%%%%%%%%%%%%%%%%%%%%%%%%%%%%%%%%%%%%%%%%%%%%%%%%%%%%%
%%%%%%%%%%%%%%%%%%%%%%%%%%%%%%%%%%%%%%%%%%%%%%%%%%%%%%%%%%%%%%%%%%%%%%%%%%%%%%%%%%%%%%%%%%%%

\begin{figure}[b!]
\begin{minipage}{0.49\textwidth}
  \begin{center}
    \includegraphics[width=0.77\textwidth]{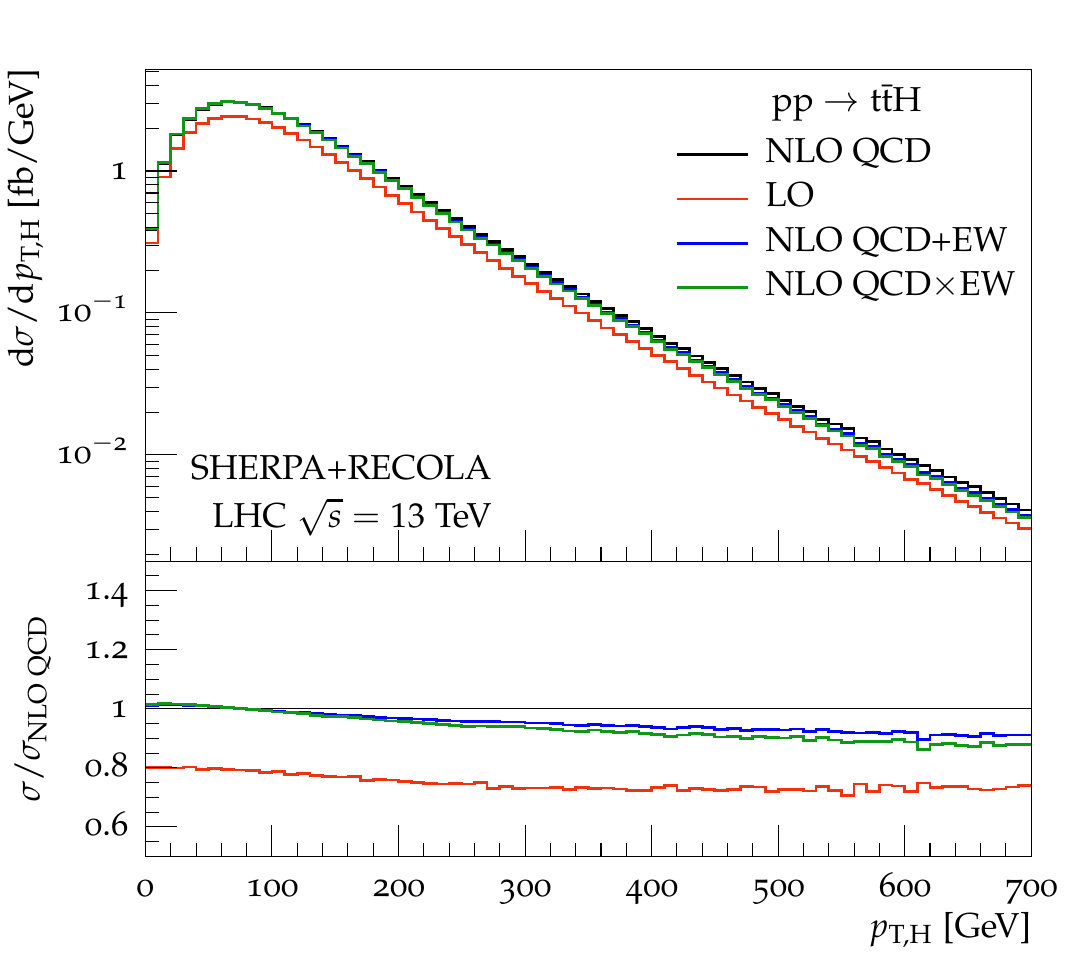}
  \end{center}
\end{minipage}
\begin{minipage}{0.49\textwidth}
  \begin{center}
    \includegraphics[width=0.77\textwidth]{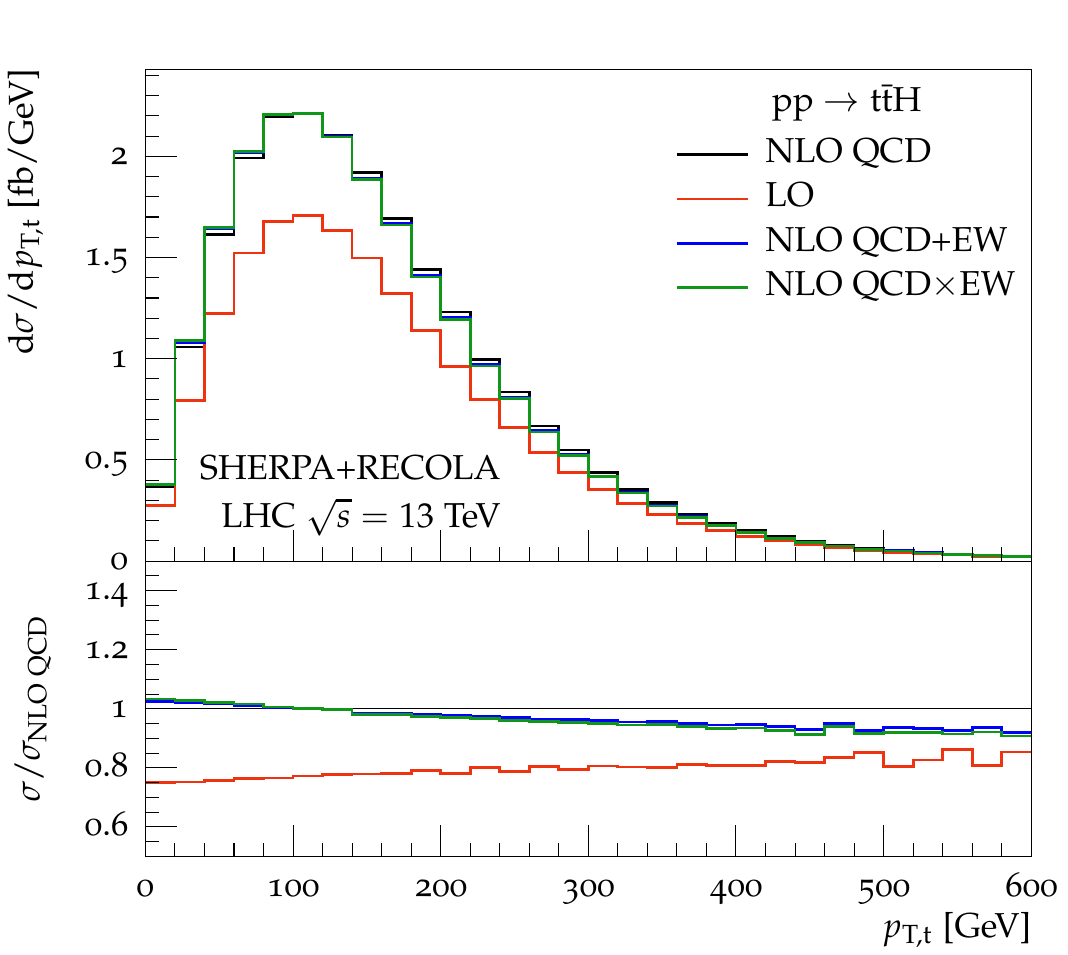}
  \end{center}
\end{minipage}
\caption{
  Transverse momentum distribution of the Higgs-boson (left panel) and the top-quark
  (right panel) in $pp\to t\bar{t}H$ at the LHC. Figures taken from Ref.~\cite{Biedermann:2017yoi}.}
\label{fig:tth}
\end{figure}

%%%%%%%%%%%%%%%%%%%%%%%%%%%%%%%%%%%%%%%%%%%%%%%%%%%%%%%%%%%%%%%%%%%%%%%%%%%%%%%%%%%%%%%%%%%%
%%%%%%%%%%%%%%%%%%%%%%%%%%%%%%%%%%%%%%%%%%%%%%%%%%%%%%%%%%%%%%%%%%%%%%%%%%%%%%%%%%%%%%%%%%%%

As an illustration, in Fig.~\ref{fig:tth} NLO QCD and EW predictions for the
process $pp\to t\bar{t}H$ evaluated with \Sherpa+\Recola~\cite{Biedermann:2017yoi} are
shown. While the dominant corrections are of QCD type, in particular for high transverse
momenta the EW corrections become sizable. The impact of yet higher-order EW corrections
can be estimated by comparing their combination with the QCD result in additive and
multiplicative manner. In particular for the top-quark transverse momentum the two variants
agree nicely, indicating factorization of the EW corrections.

%%%%%%%%%%%%%%%%%%%%%%%%%%%%%%%%%%%%%%%%%%%%%%%%%%%%%%%%%%%%%%%%%%%%%%%%%%%

\section{Matrix Elements and Parton Showers}

To connect fixed-order parton-level QCD calculations with observable particle-level
final states, parton-shower simulations prove to be an indispensable tool. Accounting
for subsequent soft/collinear emission, they in turn resum logarithms of the ordering
parameter guiding the evolution. In particular the combination of NLO QCD calculations,
accounting for the ${\cal{O}}(\alpha_s)$ corrections already, requires non-trivial matching
prescriptions, in order to preserve the NLO QCD accuracy for the inclusive production process.
Furthermore, supplementing multi-parton processes with parton showers raises the
problem of properly populating the emission phase space and consistenly setting
the multitude of scales appearing in the calculation. Furthermore, to properly account for
the significant NLO QCD corrections observed for final states with increasing parton multiplicity
necessitates means to consistently combine varying multiplicity contributions to an inclusive
production process, {\it i.e.} merging algorithms.

%%%%%%%%%%%%%%%%%%%%%%%%%%%%%%%%%%%%%%%%%%%%%%%%%%%%%%%%%%%%%%%%%%%%%%%%%%%%%%%%%%%%%%%%%%%%
%%%%%%%%%%%%%%%%%%%%%%%%%%%%%%%%%%%%%%%%%%%%%%%%%%%%%%%%%%%%%%%%%%%%%%%%%%%%%%%%%%%%%%%%%%%%

\begin{figure}[]
\begin{minipage}{0.49\textwidth}
  \begin{center}
    \includegraphics[width=0.82\textwidth]{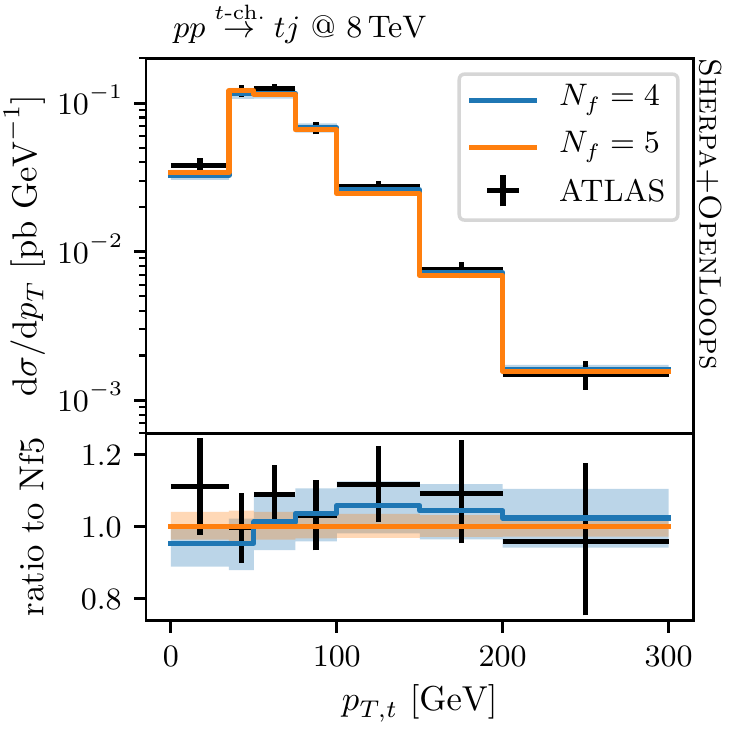}
  \end{center}
\end{minipage}
\begin{minipage}{0.49\textwidth}
  \begin{center}
    \includegraphics[width=0.82\textwidth]{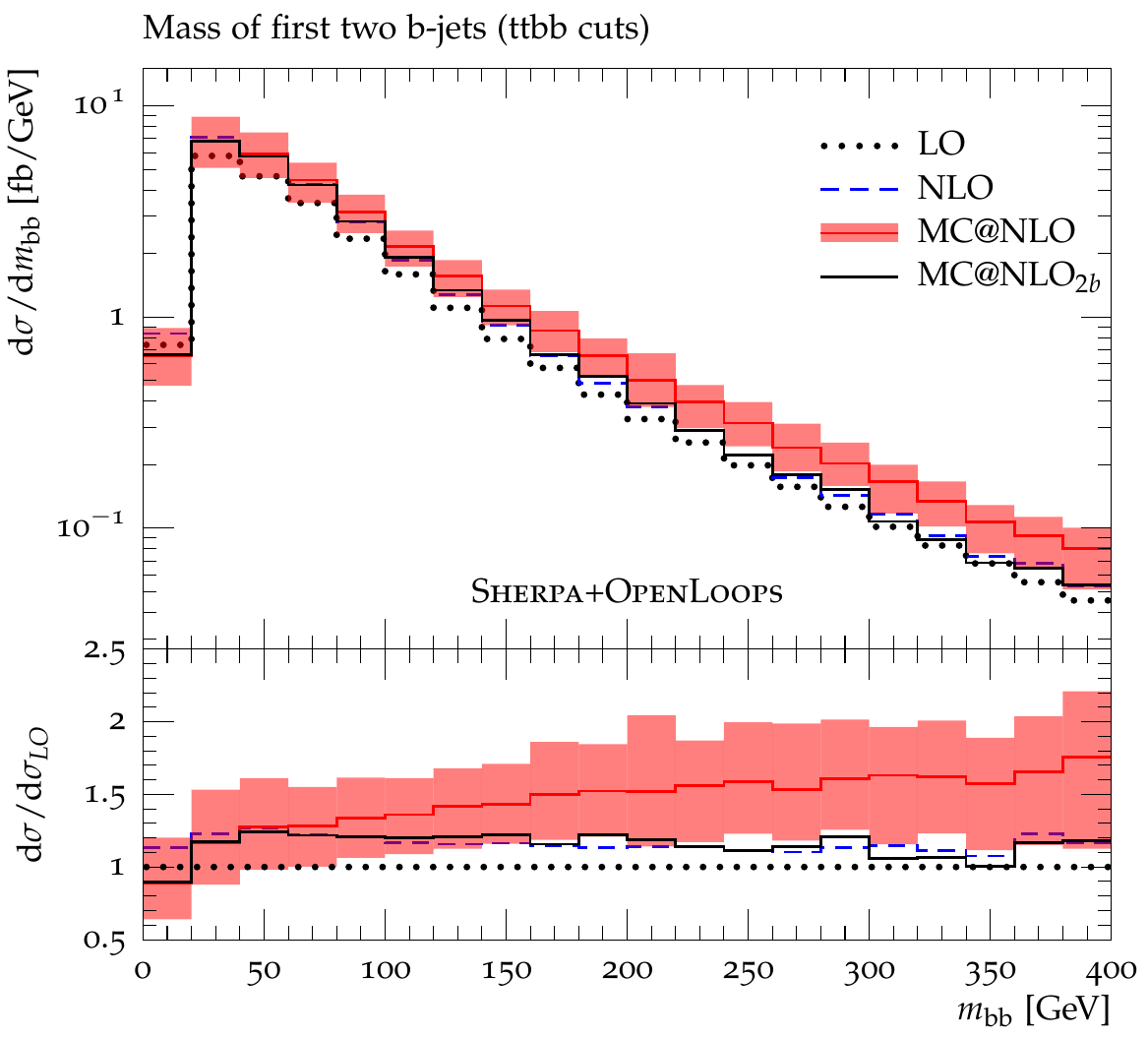}
  \end{center}
\end{minipage}
\caption{
  \Sherpa 4- and 5-flavour predictions for the top-quark transverse momentum distribution
  in $t$-channel single top production (left panel). Figure taken from Ref.~\cite{Bothmann:2017jfv}.
  Fixed-order and S-MC@NLO predictions for the two $b$-jet invariant mass in $t\bar{t}b\bar{b}$
  production (right panel). Figure taken from Ref.~\cite{Cascioli:2013era}.
}
\label{fig:tops_PS}
\end{figure}

%%%%%%%%%%%%%%%%%%%%%%%%%%%%%%%%%%%%%%%%%%%%%%%%%%%%%%%%%%%%%%%%%%%%%%%%%%%%%%%%%%%%%%%%%%%%
%%%%%%%%%%%%%%%%%%%%%%%%%%%%%%%%%%%%%%%%%%%%%%%%%%%%%%%%%%%%%%%%%%%%%%%%%%%%%%%%%%%%%%%%%%%%

In Fig.~\ref{fig:tops_PS} recent examples of attaching the \Sherpa dipole parton shower
\cite{Schumann:2007mg} to NLO QCD matrix elements are presented. In the left panel S-MC@NLO
simulations of $t$-channel single top production in the 4- and 5-flavour
scheme~\cite{Krauss:2016orf} are compared to ATLAS data~\cite{Aaboud:2017pdi}. The corresponding
uncertainty bands correspond to scale and PDF variations, obtained using an on-the-fly event
reweighting \cite{Bothmann:2016nao}. In the right panel fixed-order QCD and S-MC@NLO predictions
for the two $b$-jet invariant-mass distribution in $pp\to t\bar{t}b\bar{b}$ production are shown.
The considered core process corresponds to $t\bar{t}b\bar{b}$ matrix elements. By invoking the
parton shower a significant alteration of the $m_{bb}$ distribution is observed, having its
source in the production of additional $b$-quarks in the shower evolution \cite{Cascioli:2013era}. 

%%%%%%%%%%%%%%%%%%%%%%%%%%%%%%%%%%%%%%%%%%%%%%%%%%%%%%%%%%%%%%%%%%%%%%%%%%%%%%%%%%%%%%%%%%%%
%%%%%%%%%%%%%%%%%%%%%%%%%%%%%%%%%%%%%%%%%%%%%%%%%%%%%%%%%%%%%%%%%%%%%%%%%%%%%%%%%%%%%%%%%%%%

\begin{figure}[]
\begin{minipage}{0.49\textwidth}
  \begin{center}
    \includegraphics[width=0.93\textwidth]{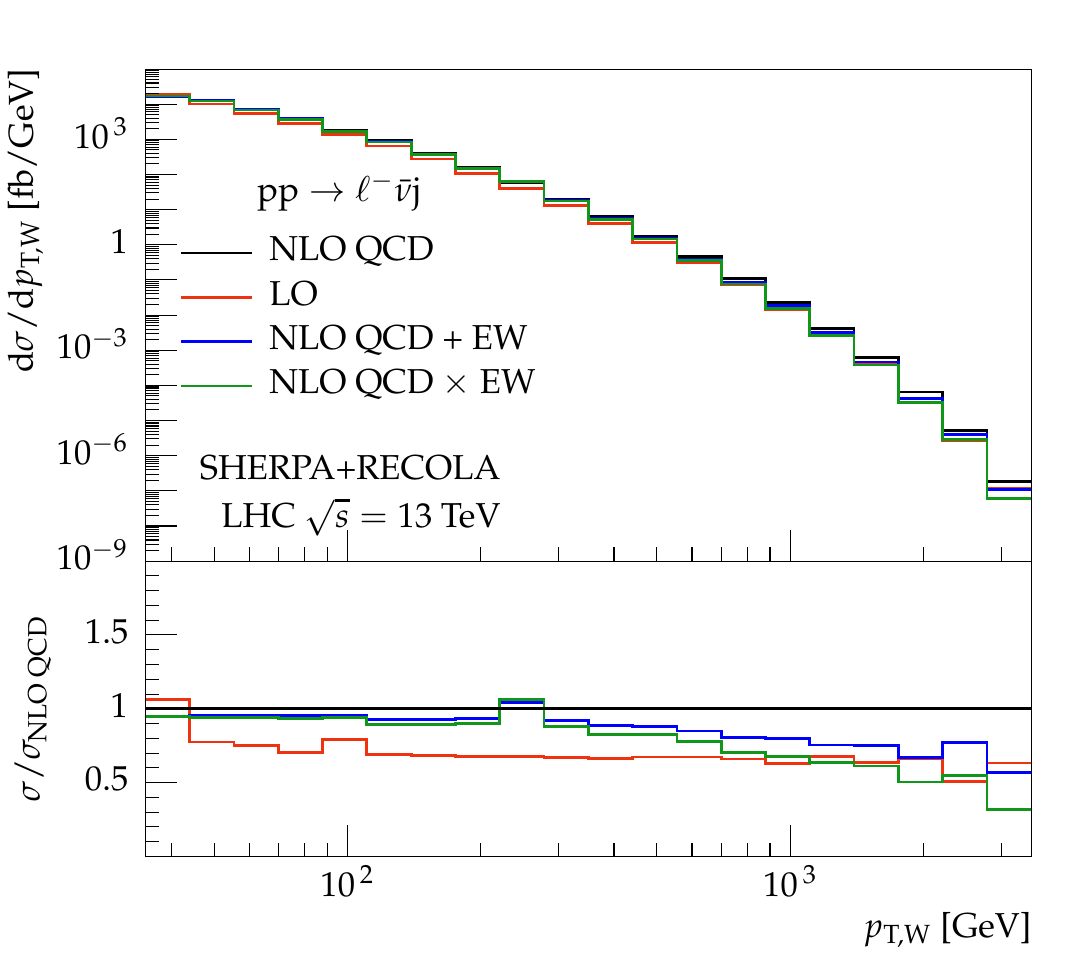}
  \end{center}
\end{minipage}
\begin{minipage}{0.49\textwidth}
  \begin{center}
    \includegraphics[width=0.7\textwidth]{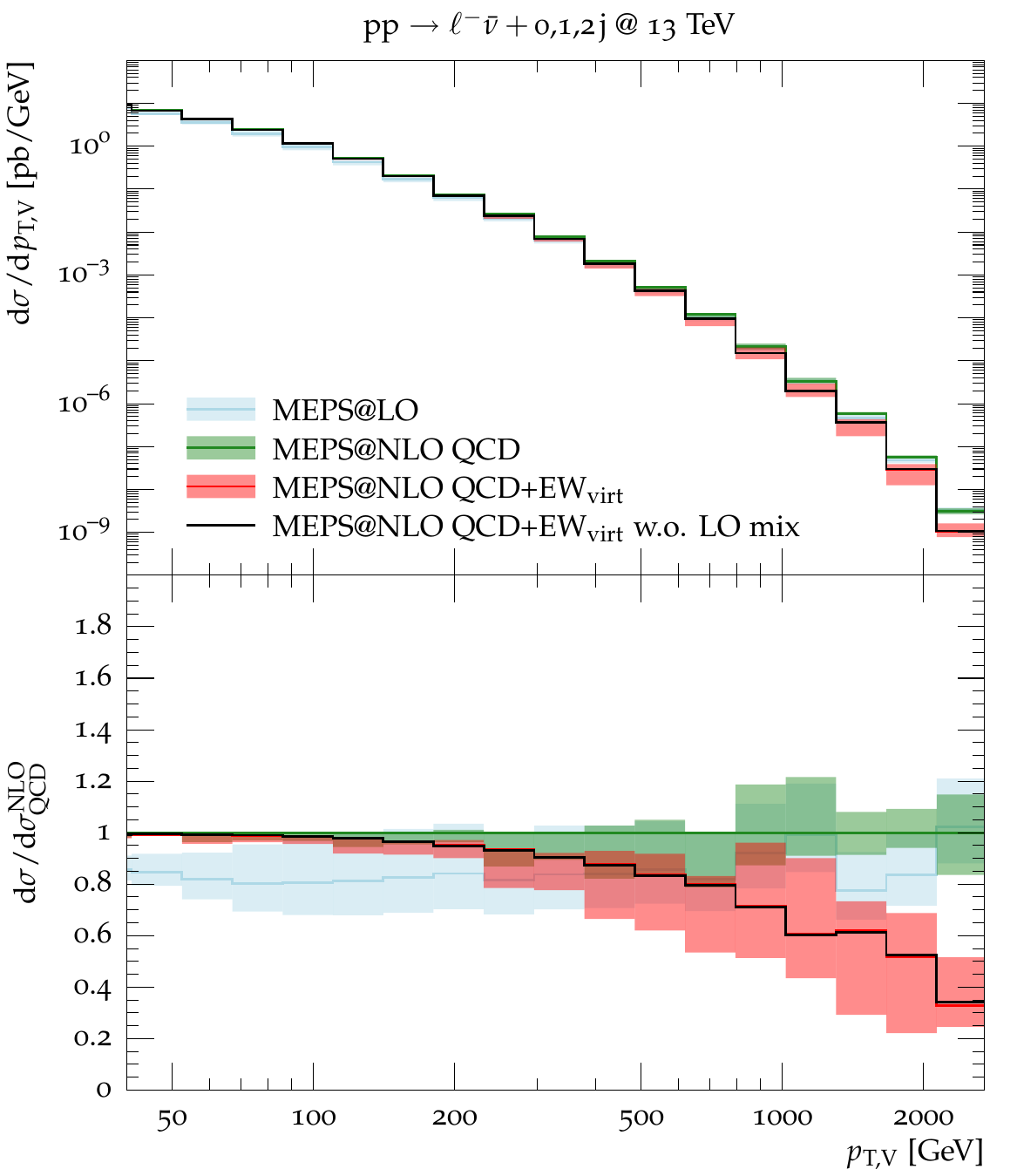}
  \end{center}
\end{minipage}
\caption{
  Fixed-order QCD and EW predictions for the gauge-boson $p_T$ in $pp\to l^-\bar{\nu}+$jet production
  (left panel). Figure taken from Ref.~\cite{Bothmann:2017jfv}. Reconstructed transverse-momentum
  distribution of the gauge boson for various matrix element plus parton shower simulations, with and
  without considering EW corrections in the virtual approximation (right panel). Figure taken from
  Ref.~\cite{Kallweit:2015dum}.}
\label{fig:Wjets_EW}
\end{figure}

%%%%%%%%%%%%%%%%%%%%%%%%%%%%%%%%%%%%%%%%%%%%%%%%%%%%%%%%%%%%%%%%%%%%%%%%%%%%%%%%%%%%%%%%%%%%
%%%%%%%%%%%%%%%%%%%%%%%%%%%%%%%%%%%%%%%%%%%%%%%%%%%%%%%%%%%%%%%%%%%%%%%%%%%%%%%%%%%%%%%%%%%%

As exemplified in Fig.~\ref{fig:tth}, EW correction often become sizable in
phase-space regions of large transverse momenta. This is furthermore illustrated in
the left panel of Fig.~\ref{fig:Wjets_EW} for the process $pp\to W^-j \to l^-\bar{\nu}j$. However,
in these regions of phase space, {\it e.g.} at high-$p_T$ of the gauge boson in $pp\to W^-j$
production, QCD corrections from additional hard emissions are significant as well and should
be considered in a realistic simulation. In the case of QCD corrections only, this is achieved
through a merging prescription \cite{Hoeche:2009rj}. However, a generalization of these formalisms
to NLO EW corrections is not yet available. Alternatively, EW corrections can be implemented in a
QCD merging algorithm in an approximated manner. In Ref.~\cite{Kallweit:2015dum} the approximation
of just considering the virtual EW correction, dubbed ${\rm EW}_{\rm virt}$, has been studied for
inclusive vector-boson production. This allows to include them as event-wise $K$-factors
in the evaluation of the various Born-level contributions. In the right panel of
Fig.~\ref{fig:Wjets_EW} corresponding predictions for merged samples with and without
the ${\rm EW}_{\rm virt}$ contributions are compared. This approximation allows for
particle-level predictions that include the full NLO QCD and the one-loop EW corrections
for the various multiplicity processes contributing to inclusive vector-boson production.

%%%%%%%%%%%%%%%%%%%%%%%%%%%%%%%%%%%%%%%%%%%%%%%%%%%%%%%%%%%%%%%%%%%%%%%%%%%%%%%%%%%%%%%%%%%%

\section{Conclusions}

Recent progress in the automated evaluation of higher-order QCD and EW corrections
in the \Sherpa generator framework have been presented. Particular emphasis was given
to top-quark production processes at the LHC. Future developments will focus on 
improvements of the parton showers and the means to consistently combined them with
NNLO QCD and NLO QCD and EW corrections.

%%%%%%%%%%%%%%%%%%%%%%%%%%%%%%%%%%%%%%%%%%%%%%%%%%%%%%%%%%%%%%%%%%%%%%%%%%%%%%%%%%%%%%%%%%%%

%\Acknowledgements
%I would like to thank the organisers of Top2017 for the kind invitation
%and for creating a very pleasant and constructive atmosphere at the conference. 
%Furthermore, I am grateful to my colleagues from the \Sherpa collaboration as
%well as Mathieu Pellen, Bendikt Biedermann and Ansgar Denner. 


\begin{thebibliography}{99}

%%
%%  bibliographic items can be constructed using the LaTeX format in SPIRES:
%%    see    http://www.slac.stanford.edu/spires/hep/latex.html
%%  SPIRES will also supply the CITATION line information; please include it.
  %%

% MC generators  
\bibitem{Buckley:2011ms}
  A.~Buckley {\it et al.},
  %``General-purpose event generators for LHC physics'',
  Phys.\ Rept.\  {\bf 504} (2011) 145.
  %doi:10.1016/j.physrep.2011.03.005
  %[arXiv:1101.2599 [hep-ph]].
  %%CITATION = doi:10.1016/j.physrep.2011.03.005;%%

%Sherpa
\bibitem{Gleisberg:2008ta}
  T.~Gleisberg {\it et al.}, 
  %``Event generation with SHERPA 1.1'',
  JHEP {\bf 0902} (2009) 007.
  %doi:10.1088/1126-6708/2009/02/007
  %[arXiv:0811.4622 [hep-ph]].
  %%CITATION = doi:10.1088/1126-6708/2009/02/007;%%

% dipole subtraction
\bibitem{Gleisberg:2007md}
  T.~Gleisberg and F.~Krauss,
  %``Automating dipole subtraction for QCD NLO calculations'',
  Eur.\ Phys.\ J.\ C {\bf 53} (2008) 501.
  %doi:10.1140/epjc/s10052-007-0495-0
  %[arXiv:0709.2881 [hep-ph]].
  
%OpenLoops
\bibitem{Cascioli:2011va}
  F.~Cascioli, P.~Maierh{\"o}fer and S.~Pozzorini,
  %``Scattering Amplitudes with Open Loops'',
  Phys.\ Rev.\ Lett.\  {\bf 108} (2012) 111601.
  %doi:10.1103/PhysRevLett.108.111601
  %[arXiv:1111.5206 [hep-ph]].
  %%CITATION = doi:10.1103/PhysRevLett.108.111601;%%

% recola  
\bibitem{Actis:2016mpe}
  S.~Actis {\it et al.}, 
 %``RECOLA: REcursive Computation of One-Loop Amplitudes'',
  Comput.\ Phys.\ Commun.\  {\bf 214} (2017) 140.
  %doi:10.1016/j.cpc.2017.01.004
  %[arXiv:1605.01090 [hep-ph]].
  %%CITATION = doi:10.1016/j.cpc.2017.01.004;%%

%tt+jets NLO QCD
\bibitem{Hoche:2016elu}
  S.~H{\"o}che {\it et al.}, 
  %``Next-to-leading order QCD predictions for top-quark pair production with up to three jets'',
  Eur.\ Phys.\ J.\ C {\bf 77} (2017) no.3,  145.
  %doi:10.1140/epjc/s10052-017-4715-y
  %[arXiv:1607.06934 [hep-ph]].
  %%CITATION = doi:10.1140/epjc/s10052-017-4715-y;%%

%recola+sherpa  
\bibitem{Biedermann:2017yoi}
  B.~Biedermann {\it et al.}, 
  %``Automation of NLO QCD and EW corrections with Sherpa and Recola'',
  Eur.\ Phys.\ J.\ C {\bf 77} (2017) 492.
  %doi:10.1140/epjc/s10052-017-5054-8
  %[arXiv:1704.05783 [hep-ph]].
  %%CITATION = doi:10.1140/epjc/s10052-017-5054-8;%%

%CS shower
\bibitem{Schumann:2007mg}
  S.~Schumann and F.~Krauss,
  %``A Parton shower algorithm based on Catani-Seymour dipole factorisation'',
  JHEP {\bf 0803} (2008) 038.
  %doi:10.1088/1126-6708/2008/03/038
  %[arXiv:0709.1027 [hep-ph]].
  %%CITATION = doi:10.1088/1126-6708/2008/03/038;%%

% flavour schemes
\bibitem{Krauss:2016orf}
  F.~Krauss, D.~Napoletano, S.~Schumann,
  %``Simulating $b$-associated production of $Z$ and Higgs bosons with the SHERPA event generator'',
  Phys.\ Rev.\ D {\bf 95} (2017) no.3,  036012.
  %doi:10.1103/PhysRevD.95.036012
  %[arXiv:1612.04640 [hep-ph]].

% ATLAS single-top
\bibitem{Aaboud:2017pdi}
  M.~Aaboud {\it et al.} [ATLAS Collaboration],
  %``Fiducial, total and differential cross-section measurements of $t$-channel single top-quark production in $pp$ collisions at 8 TeV using data collected by the ATLAS detector'',
  Eur.\ Phys.\ J.\ C {\bf 77} (2017) no.8,  531.
  %doi:10.1140/epjc/s10052-017-5061-9
  %[arXiv:1702.02859 [hep-ex]].

%BSM & decays   
%\bibitem{Hoche:2014kca}
%  S.~H{\"o}che {\it et al.}, 
%  %``Beyond Standard Model calculations with Sherpa'',
%  Eur.\ Phys.\ J.\ C {\bf 75} (2015) no.3,  135.
%  %doi:10.1140/epjc/s10052-015-3338-4
%  %[arXiv:1412.6478 [hep-ph]].
%  %%CITATION = doi:10.1140/epjc/s10052-015-3338-4;%%
    
%reweighting  
\bibitem{Bothmann:2016nao}
  E.~Bothmann, M.~Sch{\"o}nherr, S.~Schumann,
  %``Reweighting QCD matrix-element and parton-shower calculations'',
  Eur.\ Phys.\ J.\ C {\bf 76} (2016) no.11,  590.
  %doi:10.1140/epjc/s10052-016-4430-0
  %[arXiv:1606.08753 [hep-ph]].
  %%CITATION = doi:10.1140/epjc/s10052-016-4430-0;%%

%single top  
\bibitem{Bothmann:2017jfv}
  E.~Bothmann, F.~Krauss, M.~Sch{\"o}nherr,
  %``Single top-quark production with SHERPA'',
  arXiv:1711.02568 [hep-ph].
  %%CITATION = ARXIV:1711.02568;%%

% ttbb+PS  
\bibitem{Cascioli:2013era}
  F.~Cascioli {\it et al.}, 
  %``NLO matching for $t\bar t b \bar b$ production with massive $b$-quarks'',
  Phys.\ Lett.\ B {\bf 734} (2014) 210.
  %doi:10.1016/j.physletb.2014.05.040
  %[arXiv:1309.5912 [hep-ph]].
  %%CITATION = doi:10.1016/j.physletb.2014.05.040;%%

% V+jets EW
\bibitem{Kallweit:2015dum}
  S.~Kallweit {\it et al.}, 
  %``NLO QCD+EW predictions for V + jets including off-shell vector-boson decays and multijet merging'',
  JHEP {\bf 1604} (2016) 021.
  %doi:10.1007/JHEP04(2016)021
  %[arXiv:1511.08692 [hep-ph]].
  %%CITATION = doi:10.1007/JHEP04(2016)021;%%

% LO merging
\bibitem{Hoeche:2009rj}
  S.~H{\"o}che {\it et al.}, 
  %``QCD matrix elements and truncated showers'',
  JHEP {\bf 0905} (2009) 053.
  %doi:10.1088/1126-6708/2009/05/053
  %[arXiv:0903.1219 [hep-ph]].
  %%CITATION = doi:10.1088/1126-6708/2009/05/053;%%

\end{thebibliography}
\end{document}